\documentclass[amsmath,amssymb]{revtex4}

\usepackage[latin1]{inputenc}
\usepackage{graphicx}% Include figure files
\usepackage{dcolumn}% Align table columns on decimal point
\usepackage{bm}% bold math
\usepackage{color}

\begin{document}

\title{An Interaction-Free Quantum Shutter.}
\author{Juan Carlos Garc\'ia-Escart\'in}
\email{juagar@tel.uva.es}
\author{Pedro Chamorro-Posada}
\affiliation{Departamento de Teor\'ia de la Se\~{n}al y Comunicaciones e Ingenier\'ia Telem\'atica. Universidad de Valladolid. Campus Miguel Delibes. Paseo del Cementerio s/n 47011 Valladolid. Spain.}
\date{\today}

\begin{abstract}
We give a proposal for the physical implementation of a quantum shutter. This shutter is a mechanism able to close the passage of light through a slit when present. As it must be a quantum mechanical object, able to show superposition, interesting quantum effects are observed. Finally, we review some applications of the quantum shutter to build quantum memories and a CNOT gate.
\end{abstract}
\maketitle
\section{Introduction}
Measurement is a fundamental part of quantum physics. A particularly interesting phenomenon is that of interaction-free measurement, or IFM. With IFM we can obtain information on the state of an object without any interaction in the classical sense \cite{Dic81}. In \cite{EV93} Elitzur and Vaidman proved that IFM allows the detection of the presence of an object using a photon, even in the cases the photon does not interact with the object. In the Elitzur-Vaidman scheme the object, a bomb, is put in one of the arms of an interferometer. Depending on the presence or absence of the object the photon presents a different state at the output. Improved schemes for this ``Quantum Interrogation'' have been presented \cite{KWH95, KWM99, RG02} showing that high detection efficiency can be achieved. Experimental results have confirmed the feasibility of these schemes \cite{KWH95, KWM99,TGK98}.

A quantum shutter is a quantum object that ideally can reflect photons directed towards it, like the one introduced in \cite{AV03}. Combined with linear optics, it can provide a quantum memory and universal quantum logic \cite{GC05}. Here we will give the physical model of a system that behaves like this theoretical object. 

Section \ref{qshutters} presents the quantum shutter. Section \ref{ifmpd} studies the basic quantum interrogation setup, previous to the analysis of a simple one slit shutter in section \ref{qshutterslit}. Section \ref{apps} shows the applications to quantum information processing with a brief explanation on how a quantum memory and a CNOT gate can be built. In section \ref{conclusion} we discuss the results. 

\section{Quantum shutters.}
\label{qshutters}
\hspace{-3ex}

This paper will give a possible implementation for a quantum shutter device based on the slit system presented in \cite{AV03}. Such a device consists of a double slit system and a shutter mechanism that can close one of the slits reflecting the photons directed to that slit. This shutter must be a quantum object able to show superposition.
\begin{figure}[h]
\begin{center}
\includegraphics{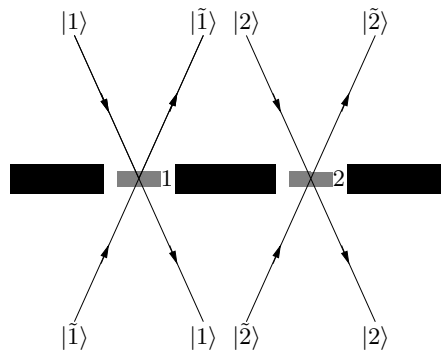}
\caption{Double slit with a quantum shutter.}
\label{shutter}
\end{center} 
\end{figure}

We will denote with $|1\rangle$ the state of a photon heading to the top port of slit number one and with $|2\rangle$ the state of a photon heading from above to the second slit.  We will call  $|\tilde{1}\rangle$ and $|\tilde{2}\rangle$ the resulting states of the photons that are reflected from the shutter when the corresponding slit is blocked. Those states correspond to the modes of the new physical ports the photons are in. If a photon goes through the slit, it will keep its state ($|1\rangle$ or $|2\rangle$). If it is reflected, the output state is $|\tilde{1}\rangle$ or $|\tilde{2}\rangle$, depending on the number of the input port (Fig. \ref{shutter}). 

Table \ref{evolutionorig} summarizes the evolution of the system after the interaction. A photon incident on the bottom side of the device with the same angle will be also in the state $|\tilde{1}\rangle$ or $|\tilde{2}\rangle$, as it will, after passing, follow the same path as a photon heading to the top side that bounces back from the shutter. If a photon incident from the bottom is reflected, it will follow the same path as the corresponding photon incident from the top that goes through the slit undisturbed.
\begin{table}[ht]
\begin{center}
\begin{tabular}{ccc}	
Input state&& Output state\\
\hline
$|\psi\rangle_{sh}=|1\rangle$&&\\
$|1\rangle$ & $\rightarrow$ & $|\tilde{1}\rangle$\\
$|2\rangle$ & $\rightarrow$ & $|2\rangle$\\
$|\tilde{1}\rangle$ & $\rightarrow$ & $|1\rangle$\\
$|\tilde{2}\rangle$ & $\rightarrow$ & $|\tilde{2}\rangle$\\
\hline
$|\psi\rangle_{sh}=|2\rangle$ &&\\
$|1\rangle$ & $\rightarrow$ & $|1\rangle$\\
$|2\rangle$ & $\rightarrow$ & $|\tilde{2}\rangle$\\
$|\tilde{1}\rangle$ & $\rightarrow$ & $|\tilde{1}\rangle$\\
$|\tilde{2}\rangle$ & $\rightarrow$ & $|2\rangle$\\
\hline
\end{tabular}
\caption{Evolution of the shutter-photon system.}
\label{evolutionorig}
\end{center}
\end{table}\\

\section{Interaction-free measurement for particle detection}
\label{ifmpd}
\begin{figure}[ht!]
\centering
\includegraphics{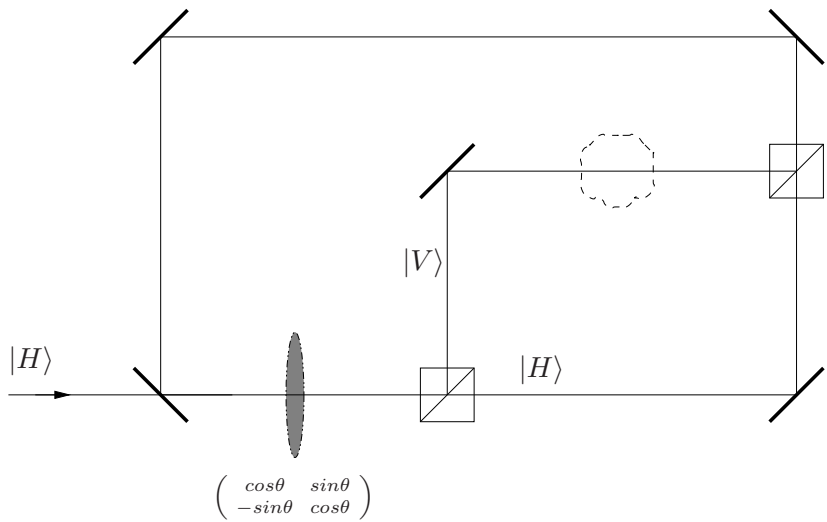}
\caption{Interaction free bomb detection system.}
\label{bomb}
\end{figure}

Before studying the quantum shutter system proper, it is useful to see a simpler case of quantum interrogation, following the proposal of \cite{KWM99}. Let's take a interferometer system like the one in figure \ref{bomb}. The particle (the bomb) can be present or not. 

We suppose that the original input is a horizontally polarized photon. The Hilbert space of the system will be given by the $\{|H\rangle,|V\rangle\}$ basis. In matrix representation
\begin{equation}
|H\rangle= \left( \begin{array}{c} 1\\ 0 \end{array} \right) , \hspace{2ex} |V\rangle= \left( \begin{array}{c} 0\\ 1 \end{array} \right).
\end{equation}

 The system has and two polarizing beamsplitters, PBS, that reflect vertically polarized photons while allowing the passage of horizontally polarized ones. The oval represents a polarization rotator, that gives the transformation $\vartheta= \left( \begin{array}{cc}
cos \theta & sin \theta \\
-sin \theta & cos \theta
\end{array} \right)$.

The interferometer has at its input the state $cos\theta|H\rangle+sin\theta|V\rangle$. If there is a bomb the probability of explosion is \'a $sin^2\theta$, and the probability it doesn't explode is $cos^2 \theta$. In the latter case the state is reduced to $|H\rangle$ and the process starts again. If the photon undergoes $N$ cycles inside the interferometer the total probability of explosion becomes $sin^{2N}\theta$, and we have a probability of $cos^{2N}\theta$ of having $|H\rangle$ as the output state. 

In the case with no bomb the superposition of states must be taken into account. If the photon goes through the circuit $N$ times the global effect will be given by the operator $\vartheta^N$. This operator can be obtained from the eigenvalues of $\vartheta$. The analysis will be useful in the following study of the shutter system. 
\begin{equation}
\left| \begin{array}{cc}
cos \theta-\lambda & sin \theta \\
-sin \theta & cos \theta-\lambda
\end{array} \right|=(cos\theta-\lambda)^2+sin^2\theta=cos^2\theta-2cos\theta\lambda+\lambda^2+sin^2\theta=\lambda^2-2cos\theta\lambda+1=0
\end{equation}
From the characteristic equation $\lambda=\frac{2cos\theta\pm\sqrt{4cos^2\theta-4}}{2}=e^{\pm i\theta}$. The eigenvectors are $\frac{1}{\sqrt{2}}\left( \begin{array}{c} 1\\ i \end{array} \right)$ associated to $e^{i\theta}$, and $\frac{1}{\sqrt{2}} \left( \begin{array}{c} 1\\ -i \end{array} \right)$ associated to $e^{-i\theta}$. Therefore,
\begin{equation}
\vartheta=e^{i\theta}\frac{1}{2}\left( \begin{array}{c} 1\\ i \end{array} \right)\left( \begin{array}{cc} 1& -i \end{array} \right)+e^{-i\theta}\frac{1}{2}\left( \begin{array}{c} 1\\ -i \end{array} \right)\left( \begin{array}{cc} 1 &i \end{array} \right)=e^{i\theta}\frac{1}{2}\left( \begin{array}{cc} 1&-i\\ i&1 \end{array} \right)+e^{-i\theta}\frac{1}{2}\left( \begin{array}{cc} 1&i\\ -i&1 \end{array} \right),
\end{equation}
and, after $N$ cycles,
\begin{equation}
\vartheta^N=e^{iN\theta}\frac{1}{2}\left( \begin{array}{c} 1\\ i \end{array} \right)\left( \begin{array}{cc} 1& -i \end{array} \right)+e^{-iN\theta}\frac{1}{2}\left( \begin{array}{c} 1\\ -i \end{array} \right)\left( \begin{array}{cc} 1 &i \end{array} \right)=\left( \begin{array}{cc} \frac{e^{iN\theta}+e^{-iN\theta}}{2}& \frac{e^{iN\theta}-e^{-iN\theta}}{2i}\\ -\frac{e^{iN\theta}-e^{-iN\theta}}{2i} &  \frac{e^{iN\theta}+e^{-iN\theta}}{2}\end{array} \right)= \left( \begin{array}{cc} 
cos (N\theta) & sin(N \theta) \\
-sin (N\theta) & cos (N\theta)
\end{array} \right).
\end{equation}

The initial state $|H\rangle$ gives the state $cos(N\theta)|H\rangle+sin(N\theta)|V\rangle$ at the output. If $\theta=\frac{\pi}{2N}$, the output becomes $|V\rangle$. When $N\rightarrow \infty$ the probability of explosion (if there is a bomb present) tends to 0. For $N=24$ the probability of explosion is less than 10\%.

\section{Quantum Shutter. Example for one slit.}
\label{qshutterslit}
\begin{figure}[ht!]
\centering
\includegraphics{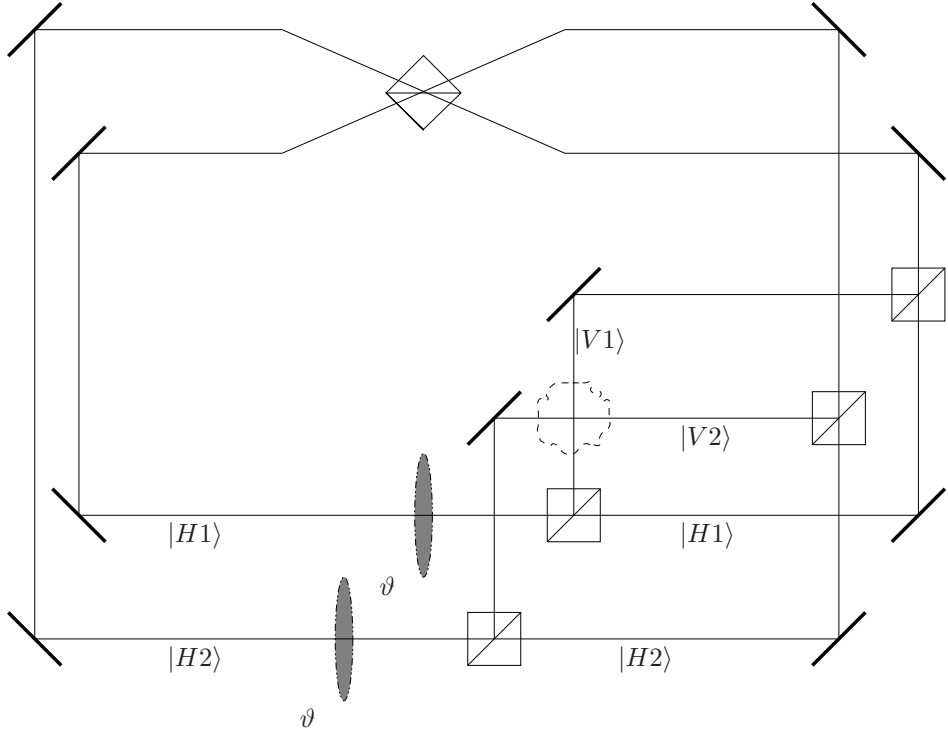}
\caption{Quantum shutter for one slit.}
\label{qshutter}
\end{figure}

We will analyze the scheme of figure \ref{qshutter}.

The system consists of two nested interferometers with the same configuration of figure \ref{bomb}, that share the same object, the bomb. Their outputs are connected by means of a PBS, so that vertically polarized photons are reflected and stay in the same interferometer and horizontally polarized photons switch their paths and get into the opposite interferometer.

The Hilbert space is given by the $\{|H1\rangle,|V1\rangle,|H2\rangle,|V2\rangle\}$ basis. Now, the photon can be in two different positions (lines) and in two different states of polarization. In matrix representation,
\begin{equation}
|H1\rangle= \left( \begin{array}{c} 1\\ 0\\0\\0 \end{array} \right) , \hspace{2ex} |V1\rangle= \left( \begin{array}{c} 0\\ 1\\0\\0 \end{array} \right) \hspace{2ex} |H2\rangle=\left( \begin{array}{c} 0\\ 0\\1\\0 \end{array} \right) , \hspace{2ex} |V2\rangle= \left( \begin{array}{c} 0\\ 0\\0\\1 \end{array} \right).
\end{equation}

The evolution after one cycle can be seen dividing the path of the photon in two parts. Before reaching the upper PBS we have two separate systems and the evolution is given by, 
\begin{equation}
U_1= \left( \begin{array}{cccc}
cos \theta & sin \theta & 0 &0\\
-sin \theta & cos \theta&0&0\\
0&0&cos \theta & sin \theta\\
0&0&-sin \theta & cos \theta
\end{array} \right).
\end{equation}

The effect of the PBS can be seen as a permutation of the first and third position of the state vector. The global effect is,
\begin{equation}
U= \left( \begin{array}{cccc}
0 & 0 & 1 &0\\
0& 1&0&0\\
1&0&0& 0\\
0&0&0&1\end{array} \right) \left( \begin{array}{cccc}
cos \theta & sin \theta & 0 &0\\
-sin \theta & cos \theta&0&0\\
0&0&cos \theta & sin \theta\\
0&0&-sin \theta & cos \theta
\end{array} \right)= \left( \begin{array}{cccc}
0&0&cos \theta & sin \theta \\
-sin \theta & cos \theta&0&0\\
cos \theta & sin \theta&0&0\\
0&0&-sin \theta & cos \theta
\end{array} \right).
\end{equation}
The eigenvalues are somewhat more difficult to find, and so are the eigenvectors, but we don't need to obtain them explicitly. Some of the values can be deduced form physical arguments. Any symbolic calculus software, such as Maple, will give us the eigenvalues $e^{i\theta}$, $e^{-i\theta}$, $1$ and $-1$. The eigenvalues $e^{\pm i \theta}$ and their corresponding eigenvectors can been easily deduced from the example in the previous section. If each of the subsystems has as its input state the eigenvector associated to $e^{i\theta}$ the state will be also preserved for the composite system. The PBS will just swap horizontally polarized photons. If they have the same probability amplitude in 1 and 2 the change will not alter the state. The same is valid for $e^{-i\theta}$. The eigenvectors must be:
\begin{equation}
e^{i\theta}\rightarrow \frac{1}{2}\left( \begin{array}{c} 1\\ i\\1\\i \end{array} \right) , \hspace{2ex} e^{-i\theta}\rightarrow \frac{1}{2}\left( \begin{array}{c} 1\\-i\\1\\-i \end{array} \right) .
\end{equation}

The eigenvalues $1$ and $-1$ were also to be expected. The evolution of the quantum state must be unitary, and from all the possible eigenvectors of modulus 1 the only angle with a physical sense was $\theta$. We don't need to know the eigenvectors that correspond to these eigenvalues.

When the particle is present, either we register an explosion, or both interforemeters end up with the photon in horizontal polarization. The PBS will take the photons from 1 to 2 and vive versa. If the initial state is $|H1\rangle$, $|H2\rangle$ or a linear combination of them, the probability of the particle absorbing the photon is $sin^2\theta$. After $N$ cycles the probability of still having the photons is $1-sin^{2N}\theta$. For an even number of cycles we recover the original state. For an odd number, the photons in $|H1\rangle$ have gone to $|H2\rangle$, and the other way round.

If there is no particle we must take into account the superposition of states. From the theorem of spectral decomposition,
\begin{equation}
U=e^{i\theta}|e_1\rangle\langle e_1|+e^{-i\theta}|e_2\rangle\langle e_2|+1|e_3\rangle\langle e_3|+(-1)|e_4\rangle\langle e_4|.
\label{desc}
\end{equation}

and $U^N$, the operator that gives the evolution after $N$ cycles, will be,
\begin{equation}
U^N=e^{iN\theta}|e_1\rangle\langle e_1|+e^{-iN\theta}|e_2\rangle\langle e_2|+1^N|e_3\rangle\langle e_3|+(-1)^{N}|e_4\rangle\langle e_4|.
\end{equation}

For an odd $N$, from (\ref{desc}),
\begin{equation}
U^N=e^{iN\theta}|e_1\rangle\langle e_1|+e^{-iN\theta}|e_2\rangle\langle e_2|+U-e^{i\theta}|e_1\rangle\langle e_1|-e^{-i\theta}|e_2\rangle\langle e_2|.
\end{equation}
\begin{equation}
U^N= \frac{1}{2}\left( \begin{array}{cccc} cos(N\theta)&sin(N\theta)&cos(N\theta)&sin(N\theta)\\ -sin(N\theta)& cos(N\theta)&-sin(N\theta)&cos(N\theta) \\ cos(N\theta)&sin(N\theta)&cos(N\theta)&sin(N\theta)\\ -sin(N\theta)& cos(N\theta)&-sin(N\theta)&cos(N\theta) \end{array} \right)+ \left( \begin{array}{cccc} 0&0&cos \theta & sin \theta \\ -sin \theta & cos \theta&0&0\\ cos \theta & sin \theta&0&0\\ 0&0&-sin \theta & cos \theta \end{array} \right)-\frac{1}{2}\left( \begin{array}{cccc} cos\theta&-sin\theta&cos\theta&-sin\theta\\ sin\theta& cos\theta&sin\theta&cos\theta \\ cos\theta&-sin\theta&cos\theta&-sin\theta\\ sin\theta& cos\theta&sin\theta&cos\theta \end{array} \right) 
\end{equation}
\begin{equation}
U^N= \frac{1}{2}\left( \begin{array}{cccc} cos(N\theta)-cos(\theta)&sin(N\theta)-sin\theta&cos(N\theta)+cos\theta&sin(N\theta)+sin\theta\\
-(sin(N\theta)+sin\theta)&cos(N\theta)+cos\theta&-(sin(N\theta)-sin\theta)&cos(N\theta)-cos(\theta)\\
cos(N\theta)+cos(\theta)&sin(N\theta)+sin\theta&cos(N\theta)-cos\theta&sin(N\theta)-sin\theta\\
-(sin(N\theta)-sin\theta)&cos(N\theta)-cos\theta&-(sin(N\theta)+sin\theta)&cos(N\theta)+cos(\theta)
\end{array} \right)
\end{equation}
Now we use the trigonometric relations, 
\begin{eqnarray}
cos(A)+cos(B)&=&2cos\left(\frac{A+B}{2}\right)cos\left(\frac{A-B}{2}\right)\\
cos(A)-cos(B)&=&-2sin\left(\frac{A+B}{2}\right)sin\left(\frac{A-B}{2}\right)\\
sin(A)+sin(B)&=&2sin\left(\frac{A+B}{2}\right)cos\left(\frac{A-B}{2}\right)\\
sin(A)-sin(B)&=&2cos\left(\frac{A+B}{2}\right)sin\left(\frac{A-B}{2}\right).
\end{eqnarray}
\begin{equation}
U^N= \left( \begin{array}{cccc} 
-sin\left(\frac{N+1}{2}\theta\right)sin\left(\frac{N-1}{2}\theta\right)&cos\left(\frac{N+1}{2}\theta\right)sin\left(\frac{N-1}{2}\theta\right)&cos\left(\frac{N+1}{2}\theta\right)cos\left(\frac{N-1}{2}\theta\right)&sin\left(\frac{N+1}{2}\theta\right)cos\left(\frac{N-1}{2}\theta\right)\\
-sin\left(\frac{N+1}{2}\theta\right)cos\left(\frac{N-1}{2}\theta\right)&cos\left(\frac{N+1}{2}\theta\right)cos\left(\frac{N-1}{2}\theta\right)&-cos\left(\frac{N+1}{2}\theta\right)sin\left(\frac{N-1}{2}\theta\right)&-sin\left(\frac{N+1}{2}\theta\right)sin\left(\frac{N-1}{2}\theta\right)\\
cos\left(\frac{N+1}{2}\theta\right)cos\left(\frac{N-1}{2}\theta\right)&sin\left(\frac{N+1}{2}\theta\right)cos\left(\frac{N-1}{2}\theta\right)&-sin\left(\frac{N+1}{2}\theta\right)sin\left(\frac{N-1}{2}\theta\right)&cos\left(\frac{N+1}{2}\theta\right)sin\left(\frac{N-1}{2}\theta\right)\\
-cos\left(\frac{N+1}{2}\theta\right)sin\left(\frac{N-1}{2}\theta\right)&-sin\left(\frac{N+1}{2}\theta\right)sin\left(\frac{N-1}{2}\theta\right)&-sin\left(\frac{N+1}{2}\theta\right)cos\left(\frac{N-1}{2}\theta\right)&cos\left(\frac{N+1}{2}\theta\right)cos\left(\frac{N-1}{2}\theta\right)
\end{array} \right)
\end{equation}

For $\theta=\frac{\pi}{N+1}$, 
\begin{equation}
U^N= \left( \begin{array}{cccc} 
-sin\left(\frac{N-1}{2}\theta\right)&0&0&cos\left(\frac{N-1}{2}\theta\right)\\
-cos\left(\frac{N-1}{2}\theta\right)&0&0&-sin\left(\frac{N-1}{2}\theta\right)\\
0&cos\left(\frac{N-1}{2}\theta\right)&-sin\left(\frac{N-1}{2}\theta\right)&0\\
0&-sin\left(\frac{N-1}{2}\theta\right)&-cos\left(\frac{N-1}{2}\theta\right)&0
\end{array} \right).
\end{equation}

Now, when $N\rightarrow \infty$ $\frac{N-1}{2}\theta=\frac{N-1}{N+1}\frac{\pi}{2}=\frac{\pi}{2}$, and we get,
\begin{equation}
U^N= \left( \begin{array}{cccc} 
-1&0&0&0\\
0&0&0&-1\\
0&0&-1&0\\
0&-1&0&1
\end{array} \right).
\end{equation}

Any input in the state $|H1\rangle$, $|H2\rangle$ or a linear combination will exit the system in the same port but with a $\pi$ phase shift. The probability of going out in a vertically polarized state can be made arbitrarily small by choosing a large enough value of $N$, as long as it is odd. In no case there will be a photon coming out of a port it didn't enter.

This corresponds to the behaviour of a quantum shutter for a single slit if we identify $|H1\rangle$ with $|1\rangle$ and $|H2\rangle$ with $|\tilde{1}\rangle$. If there is a particle the photon will change its port. If there is no particle the port is kept, with a sign shift in the state. If the particle can be in this system, that we will label as 1, or in another one identical to first one, labelled as 2, we get the operation shown in table \ref{evolution}.
\begin{table}[ht]
\begin{center}
\begin{tabular}{ccc}	
Input state&&Output\\
\hline
$|\psi\rangle_{sh}=|1\rangle$&&\\
$|1\rangle$ & $\rightarrow$ & $|\tilde{1}\rangle$\\
$|2\rangle$ & $\rightarrow$ & -$|2\rangle$\\
$|\tilde{1}\rangle$ & $\rightarrow$ & $|1\rangle$\\
$|\tilde{2}\rangle$ & $\rightarrow$ & -$|\tilde{2}\rangle$\\
\hline
$|\psi\rangle_{sh}=|2\rangle$ &&\\
$|1\rangle$ & $\rightarrow$ & -$|1\rangle$\\
$|2\rangle$ & $\rightarrow$ & $|\tilde{2}\rangle$\\
$|\tilde{1}\rangle$ & $\rightarrow$ & -$|\tilde{1}\rangle$\\
$|\tilde{2}\rangle$ & $\rightarrow$ & $|2\rangle$\\
\hline
\end{tabular}
\caption{Evolution of the shutter system.}
\label{evolution}
\end{center}
\end{table}

\section{Applications to quantum information processing}
\label{apps}
A system with a behaviour like the one of the previous section can perform all the functions of the shutter system described in \cite{GC05} providing us with a quantum memory and a CNOT gate. 
\subsection{Quantum memory.}
\label{qm}
We can store data using the correspondence $\frac{|1\rangle-|\tilde{1}\rangle}{\sqrt{2}}$ for the logical $|0\rangle$ state, and $\frac{|2\rangle+|\tilde{2}\rangle}{\sqrt{2}}$ for a logical $|1\rangle$. The shutter system will start in the $|+\rangle$ state.

\begin{equation}
|+\rangle (\alpha |0\rangle+ \beta |1\rangle)\rightarrow  |+\rangle \left( \alpha \frac{|1\rangle-|\tilde{1}\rangle}{\sqrt{2}}+\beta \frac{|2\rangle+|\tilde{2}\rangle}{\sqrt{2}} \right)
\end{equation} 

After the interaction, the state will be, 
\begin{equation}
-\alpha|+\rangle\frac{|1\rangle-|\tilde{1}\rangle}{\sqrt{2}}-\beta |-\rangle\frac{|2\rangle+|\tilde{2}\rangle}{\sqrt{2}}.
\end{equation} 

Now we have a shutter state associated to each of the logical states. We can eliminate the photon applying Hadamard gates to the ports $|2\rangle$ and $|1\rangle$, and $|\tilde{2}\rangle$ and $|\tilde{1}\rangle$, respectively:
\begin{equation}
-\alpha |+\rangle\frac{-|1\rangle+|2\rangle+|\tilde{1}\rangle-|\tilde{2}\rangle}{2}-\beta |-\rangle\frac{|1\rangle+|2\rangle+|\tilde{1}\rangle+|\tilde{2}\rangle}{2}.
\end{equation} 
When measuring the ports we put a classical bit $a$ to 0 if the photon is found in the state $|1\rangle$ or $|\tilde{2}\rangle$, and to 1 if it is found in $|2\rangle$ o $|\tilde{1}\rangle$. The shutter state will be $(-1)^a\alpha|+\rangle-\beta|-\rangle$. 

We can read the qubit directing the state $\frac{|1\rangle-|\tilde{1}\rangle+|2\rangle+|\tilde{2}\rangle}{2}$ to the shutter system. After the interaction, 
\begin{equation}
\frac{-(-1)^a\alpha |+\rangle\frac{|1\rangle-|\tilde{1}\rangle}{\sqrt{2}}-(-1)^a\alpha|-\rangle\frac{|2\rangle+|\tilde{2}\rangle}{\sqrt{2}}+\beta|-\rangle\frac{|1\rangle-|\tilde{1}\rangle}{\sqrt{2}}+\beta|+\rangle\frac{|2\rangle+|\tilde{2}\rangle}{\sqrt{2}}}{\sqrt{2}}.
\end{equation} 
If we measure the state of the shutter, for a measurement giving $|+\rangle$ we have the photon state $-(-1)^a\alpha\frac{|1\rangle-|\tilde{1}\rangle}{\beta\sqrt{2}}+\frac{|2\rangle+|\tilde{2}\rangle}{\sqrt{2}}$. If the measurent gives $|-\rangle$ we have $-(-1)^a\alpha\frac{|2\rangle+|\tilde{2}\rangle}{\beta\sqrt{2}}+\frac{|1\rangle-|\tilde{1}\rangle}{\sqrt{2}}$. If we associate a classical bit to the measurements result ($b=0$ for $|+\rangle$, $b=1$ for $|-\rangle$) the logical state will be: $-(-1)^a\alpha|b\rangle+\beta|b\oplus1\rangle$. We can recover the original qubit with a classically controlled NOT gate, controlled by $b$ and a doubly controlled (by $a$ and $|0\rangle$) sign shift.

An optical implementation for all the necessary classically controlled gates is given in \cite{GC05}.

\subsection{CNOT Gate.}
\label{CNOT}
We start from two qubits with state $\alpha|00\rangle+\beta|01\rangle+\gamma|10\rangle+\delta|11\rangle$. If these two qubits are stored in a quantum shutter memory system the state of the shutters will be $(-1)^{a_1+a_2}\alpha|++\rangle-(-1)^{a_2}\beta|+-\rangle-(-1)^{a_1}\gamma|-+\rangle+\delta|--\rangle$.

Now, we direct the state $\frac{|1\rangle-|\tilde{1}\rangle}{2}+(-1)^{a_1}\frac{|2\rangle+|\tilde{2}\rangle}{2}$ to the second qubit, the leftmost one. For the shake of clarity we will put the state of the photons by their corresponding logical state ($\frac{|1\rangle-|\tilde{1}\rangle}{\sqrt{2}}=|0\rangle$, $\frac{|2\rangle+|\tilde{2}\rangle}{\sqrt{2}}=|1\rangle$). 

The output from the second slit of the second shutter system are redirected to the slit 2 of the first shutter, so that its state is changed when in the first qubit we had a $|1\rangle$. After the interaction:
\begin{eqnarray}
\frac{1}{2}[&- |+\rangle|0\rangle((-1)^{a_1+a_2}\alpha|+\rangle-(-1)^{a_2}\beta|-\rangle)&+|-\rangle|1\rangle((-1)^{a_2}\alpha|-\rangle-(-1)^{a_1+a_2}\beta|+\rangle)\nonumber\\
&-|-\rangle|0\rangle(-(-1)^{a_1}\gamma|+\rangle+\delta|-\rangle)&+|+\rangle|1\rangle(-\gamma|+\rangle+(-1)^{a_1}\delta|-\rangle).
\end{eqnarray} 
When reading the first qubit, we can measure the shutter state. Depending on the state we find we have different photon states:
\begin{eqnarray}
&|++\rangle:& (-1)^{a_1+a_2}\alpha|00\rangle-(-1)^{a_2}\beta|01\rangle+\gamma|11\rangle-(-1)^{a_1}\delta|10\rangle.\nonumber\\
&|+-\rangle:& (-1)^{a_1+a_2}\alpha|01\rangle-(-1)^{a_2}\beta|00\rangle+\gamma|10\rangle-(-1)^{a_1}\delta|11\rangle.\nonumber\\
&|-+\rangle:& -(-1)^{a_2}\alpha|11\rangle-(-1)^{a_1+a_2}\beta|10\rangle-(-1)^{a_1}\gamma|00\rangle+\delta|01\rangle.\nonumber\\
&|--\rangle:& -(-1)^{a_2}\alpha|10\rangle-(-1)^{a_1+a_2}\beta|11\rangle-(-1)^{a_1}\gamma|01\rangle+\delta|00\rangle.
\end{eqnarray} 

In a similar fashion to \cite{GC05} we can apply a series of classically controlled optical gates to these results to recover the state $\alpha|00\rangle+\beta|01\rangle+\gamma|11\rangle+\delta|10\rangle$, which is the CNOT of the original input qubits. The gates will be different for different measured shutter states. 

\section{Conclusion}
\label{conclusion}
We have seen that a quantum interrogation setup can operate as a quantum shutter. An input photon will alter its output state depending on the presence or absence of a quantum object. The construction of this system would lead to the implementation of a quantum memory and, from it, a CNOT gate. This presents an interesting alternative to other quantum computer implementation proporsals, especially if the quantum object doesn't need to be highly coupled to a cavity, as occurs in the quantum interrogation model in \cite{GC05b}

\bibliographystyle{apsrev}
\bibliography{lspace} % Produces the bibliography via BibTeX.

\end{document}